\documentclass[aps,prl,twocolumn,groupedaddress,showpacs]{revtex4}

\usepackage{graphicx}

\begin{document}

\title{Matter-wave interferometer for large molecules}

\author{Bj\"{o}rn Brezger}
\author{Lucia Hackerm\"{u}ller}
\author{Stefan Uttenthaler}
\author{Julia Petschinka}
\author{Markus Arndt}
\author{Anton Zeilinger}
\email[]{zeilinger-office@exp.univie.ac.at}
\homepage[]{www.quantum.univie.ac.at}
\affiliation{Universit\"{a}t Wien, Institut f\"{u}r Experimentalphysik,
Boltzmanngasse 5, A-1090 Wien, Austria}

\date{January 31, 2002}

\begin{abstract} 
We demonstrate a near-field Talbot-Lau interferometer for C$_{70}$ fullerene
molecules. Such interferometers are particularly suitable for larger masses.
Using three free-standing gold gratings of 1\,$\mu$m period and a transversally
incoherent but velocity-selected molecular beam, we achieve an interference
fringe visibility of 40~\% with high count rate. Both the high visibility and
its velocity dependence are in good agreement with a quantum simulation that
takes into account the van der Waals interaction of the molecules with the
gratings and are in striking contrast to a classical moir\'{e} model.
\end{abstract}


\maketitle



Interferometry with matter waves has become a large field of interest throughout
the last years \cite{Aspect2001a,Berman1997a}. It represents a powerful tool for
the demonstration of basic quantum phenomena and of matter wave effects as well as
for applications in high precision measurements of inertial forces
\cite{Rauch2000a,Gustavson1997a} and fundamental constants \cite{Weiss1993a}.

It is therefore interesting to extend techniques from neutron and atom
interferometry to more massive, complex objects, like large molecules, in order to
quantitatively study decoherence by thermal coupling to the environment, to get
information about a wide range of molecular properties and to work on novel
lithographical or metrological applications \cite{ArndtBellbook}. In this Letter,
we report the first demonstration of an interferometer for macromolecules, and in
particular a Talbot-Lau interferometer for C$_{70}$. The molecules are in
different internal states with many excited vibrational and rotational degrees of
freedom. Nevertheless we observe a clear signature for a quantized center-of-mass
motion.

Up to now two types of interferometers with molecules have been demonstrated,
namely a Ramsey-Bord\'{e} interferometer using I$_{2}$ \cite{Borde1994a} and a
mechanical Mach-Zehnder interferometer for Na$_2$ \cite{Chapman1995b}. The use
of such devices for larger molecules remains however a challenge: A direct
application of a Ramsey-Bord\'{e} interferometer for complex molecules is
precluded by the lack of narrow resonant transitions. On the other hand, a
simple extrapolation of the Mach-Zehnder interferometer to heavy, fast molecules
would require extremely fine gratings, good collimation or large distances. This
is because such an interferometer operates in the Fraunhofer regime where the
characteristic size of a diffraction pattern scales linearly both with the
wavelength and with the distance between the diffracting structure and the plane
of observation.

 \begin{figure}
 \includegraphics[width=0.97\columnwidth]{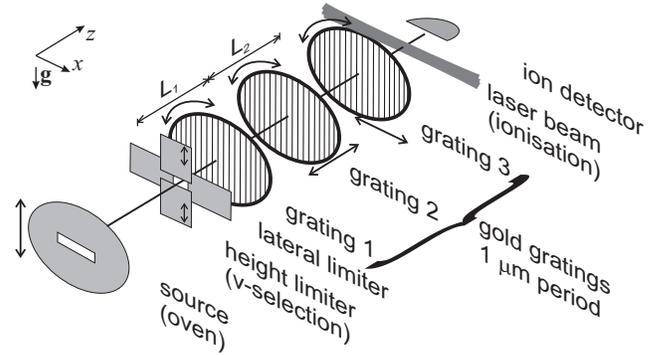}
 \caption[setup]{\label{fig:setup}
 Setup of the Talbot-Lau interferometer inside the vacuum chamber:
 It consists of three gratings which are incoherently illuminated by the
 molecule
 beam. The interference fringes are detected by transversal scanning of the third
 grating and integral detection using an ionizing laser beam. The tilt of
 each grating and the longitudinal position of the second one are
 critical and may be adjusted with actuators. The height limiter
 ensures that only the trajectories within a narrow velocity range,
 tunable by varying the vertical position of the oven, pass from the oven to the detector.
 }
 \end{figure}

A solution to this problem is the Talbot-Lau interferometer, which also consists
of three successive gratings but operates in the near-field or Fresnel regime
where the characteristic size of a diffraction pattern scales with the square root
both of the wavelength and of the distance. A Talbot-Lau interferometer can accept
a spatially incoherent beam which implies that no collimation is needed, and it
works with a spatially extended detector. Therefore Talbot-Lau count rates can
exceed those of Mach-Zehnder interferometers by several orders of magnitude.

A Talbot-Lau interferometer has already been demonstrated for potassium atoms
\cite{Clauser1994a} and -- in the time domain -- for sodium atoms in a
Bose-Einstein condensate \cite{Deng1999b}. As the scaling properties of
near-field devices are very favorable in comparison to far-field
interferometers, the Talbot-Lau effect has actually been proposed for
experiments using quantum objects up to the size of a
virus~\cite{Clauser1997a,ArndtBellbook}.

The successful diffraction of fullerenes at both a mechanical \cite{Arndt1999a}
and an optical \cite{Nairz01a} grating has now stimulated a new set of
experiments exploring the limits of macromolecule interferometry. The experiment
presented here takes advantage of the Talbot effect which is a self-imaging
phenomenon that occurs when a periodic structure is illuminated by a coherent
beam~\cite{Talbot1836,Chapman1995a,Patorski1989a}. Images of this grating are
then reconstructed at discrete multiples of the Talbot length $L_T =
{d^2}/{\lambda_{dB}}$ where $d$ is the grating constant and $\lambda_{dB}$ the
de Broglie wave length of the incident object.

Our interferometer consists of three identical gratings which are equally spaced.
The first grating acts as a comb of sources which have vanishing mutual coherence.
Yet each source is sufficiently small to prepare the required transverse coherence
at the second grating. The second grating is then coherently imaged onto the plane
of the third grating by the Talbot effect. The images coming from all sources look
identical but they are shifted by multiples of the grating constant and therefore
overlap in spite of their lack of mutual coherence \cite{Patorski1989a}. A movable
third grating blocks or transmits the molecular interference fringes according to
its position.

The experimental setup, as sketched in Fig.~\ref{fig:setup}, consists of three
free-standing gold gratings (Heidenhain) with a nominal period of $d=(991.25\pm
0.25)\,$nm, an open fraction (ratio of slit width to period) of $f=0.48\pm 0.02$,
a thickness of $b=500\,$nm and a huge usable area with a diameter of 16\,mm. The
distance between the gratings is set to $L_1=L_2=0.22\,$m. The whole setup is
placed in a vacuum chamber at a pressure of $3\cdot 10^{-8}\,$mbar. A beam of
C$_{70}$ fullerene molecules is generated by sublimation in an oven at a
temperature of 650$\,^\circ$C. Its velocity distribution is close to that of an
effusive source, with a most probable velocity of about 200\,m/s. A laser beam
with a power of 26\,W traverses the apparatus behind the third grating in
horizontal direction. It is generated by a multi-line visible argon ion laser
focussed to a beam waist of 8$\,\mu$m ($1/e^2$-radius). It ionizes the arriving
molecules regardless of their transversal position. The interferometer signal is
obtained by detecting the resulting ions and scanning the third grating
transversally, using an actively stabilized piezo translation stage (Piezosystem
Jena).

The fullerene beam travels 2.38\,m from the oven to the detection
laser. In the horizontal direction, the beam is restricted only
by the rectangular oven orifice of 1.2\,mm length and a
500\,$\mu$m wide slit that determines the used segment of the
first grating. This means that in comparison to the diffraction
angles of about 1\,$\mu$rad, the illumination is transversally
incoherent.

In the vertical direction, the oven orifice of 200\,$\mu$m height and the laser
beam fix two regions where the trajectories of the detected molecules are
tightly confined in comparison to a typical free fall distance during their
flight. Halfway in between, the vertical positions of the detected molecules are
therefore correlated with their longitudinal velocities. We introduce an
adjustable slit (Piezosystem Jena) that limits the beam height to 150\,$\mu$m or
less at a distance of 1.38\,m from the oven, 0.12\,m before the first grating.
It selects a narrow velocity distribution, which has been measured using a
mechanical chopper in front of the oven orifice. Its center may be varied
between 80 and 215 m/s by adjusting the vertical oven position. This corresponds
to de Broglie wavelengths of 5.9 and 2.2~pm, respectively. The FWHM of the
wavelength distribution goes up from 8\,\% of the mean to 35\,\% with decreasing
center wavelength.


 \begin{figure}
 \includegraphics[width=0.97\columnwidth]{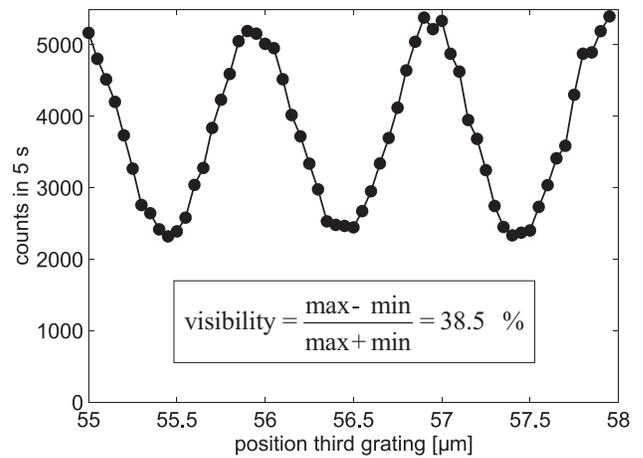}
 \caption[single scan]{\label{fig:singlescan}
 Interference fringes (raw data)
 resulting from a typical single scan of the third grating. A central
 velocity of 115\,m/s was selected.
 }
 \end{figure}

Fig.~\ref{fig:singlescan} shows the detected signal from a single scan at nearly
optimal settings for maximum contrast. Apart from some noise it corresponds well
to a sine, as expected from theory for our open fractions near 0.5, which suppress
higher Fourier components. Fast Fourier transformation (FFT) gives a sharp single
peak corresponding to the lattice period. A signal-to-noise ratio of 50:1 was
achieved in just 150\,s of total measuring time. It was not necessary to correct
the extracted visibilities for dark counts because the dark count rate was about
0.2\,s$^{-1}$ compared to count rates in the range from 50\,s$^{-1}$ to
450\,s$^{-1}$ at central velocities of 80\,m/s and 160\,m/s, respectively
\cite{footnote1}. The phase of the peak FFT component gives the spatial position
of the fringe pattern. Comparison of subsequent scans gives the lateral drift of
the three-grating setup, which is of the order of 2\,nm/min.

Tilting the three gratings with respect to each other diminishes the visibility.
Each grating was prealigned to several milliradians with respect to gravity by
observing a laser diffraction pattern outside the vacuum chamber. Fine
adjustment of the grating parallelism --- better than 2\,mrad --- was achieved
by maximizing the fullerene fringe visibility. A much smaller variation of the
visibility was observed when all three gratings were tilted by the same amount.
This is because the calculated interferometer phase shift from gravitation is
$\Delta \varphi_g=0.2\,$rad per mrad of tilt (see below) and only its
non-uniformity due to the finite velocity distribution diminishes the
visibility.

Longitudinal displacement of one grating also affects the visibility. Again, the
requirement $L_1=L_2$ was prealigned before evacuating the chamber and
fine-aligned on the 100\,$\mu$m level by maximizing the contrast. Vibration
isolation of the optical table by pneumatic feet proved to be crucial for
obtaining a high visibility.

 \begin{figure}
 \includegraphics[width=0.97\columnwidth]{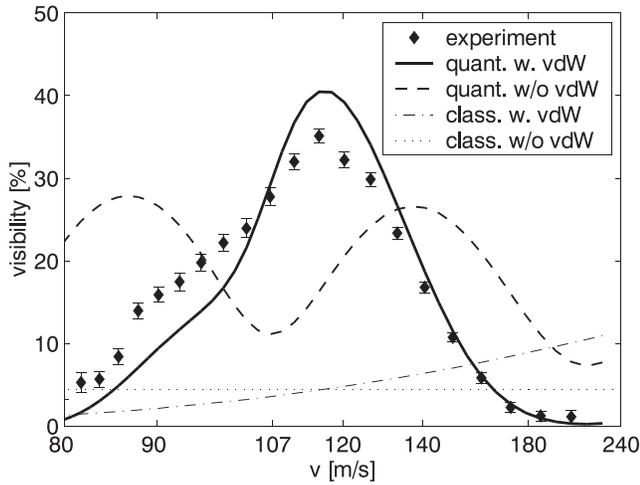}
 \caption[v dependence]{\label{fig:vdepend}
 Dependence of the interference fringe visibility on the
 mean velocity of the molecular beam.
 Numerical simulation results are plotted for
 four models without free parameters:
 classical or quantum behavior, with or without consideration of the van
 der Waals (vdW) interaction of the molecules with the second grating. The
 quantum result including the van der Waals effect is clearly the only adequate one.
}
 \end{figure}

The appearance of a periodic signal as a function of the position of the third
grating is not necessarily a sign of quantum interference --- it could also result
from classical moir\'e fringes, i.~e.\ shadow patterns which result from the
geometry of straight rays passing the gratings or being blocked at one of them
\cite{Oberthaler1996a}. These patterns do not depend on the particle velocity, in
sharp contrast to the behavior of quantum de Broglie waves: here the velocity
gives the wavelength, hence the Talbot length, and the ratio $L_1/L_T$ of the
grating separation to the Talbot length is a crucial parameter in the theory of
wave propagation.

The quantitative evaluation of the expected velocity dependence involved numerical
simulations of the three-grating setup in a quantum de Broglie wave model, in
comparison with a classical point-particle calculation. The quantum model relies
on Fresnel integrals and calculates them efficiently by a Fourier transform method
under the assumption of incoherent illumination. Assuming binary transmission
functions corresponding to our grating geometry, the classical point particle
model yields a velocity-independent fringe visibility of 5\,\%, whereas the
quantum model gives the dashed curve in Fig.~\ref{fig:vdepend} with two maxima
slightly below 30\,\% in the experimentally accessible velocity range. A relative
minimum in between corresponds to the velocity $v_T=107\,$m/s where $L_{T}=L_1$.
In this simulation and the following ones, our measured velocity distributions
have been taken into account.

The experimental visibility curve (diamonds in Fig.~\ref{fig:vdepend}) exhibits
a pronounced velocity dependence and a maximum visibility of 35\,\% in the
vicinity of the one-Talbot-length criterion. On other days, up to 39~\% were
obtained with the same settings, and 41~\% with a narrower height limiter and
therefore a narrower velocity distribution. This is totally incompatible with
classical moir\'e fringes but corresponds also quite poorly to the described
quantum simulation result.

The most likely reason for this discrepancy is that the ansatz of a purely
absorptive grating transmission function is not adequate: the outcome of far-field
diffraction experiments with rare-gas atoms \cite{Grisenti1999a} is described
correctly only by a theory that includes the phase shift by the van der Waals
interaction between the particles and the grating. To estimate the magnitude of
this effect for our experimental situation, we assumed a potential $V(r)=-C_3
r^{-3}$ for a molecule at a distance $r\gtrsim 1\,$nm from a gold surface. The
constant $C_3$ is known for various noble gases \cite{Vidali1991a} near gold
surfaces and depends roughly linearly on their atomic dc polarizability. We
extrapolated to the known dc polarizability
$\alpha_{\mathrm{DC}}=97\,$\AA$^3\times 4\pi\epsilon_0$ of the fullerene molecule
C$_{70}$ \cite{Dresselhaus1998a} and obtained the rough estimate
$C_3^{(Au-C70)}\approx 0.09\,$eV\,nm$^3$. For the passage through a grating slit
of width $f d$ ($=475\,$nm) centered at $x=0$ we obtain

\begin{equation}
\label{eq:potint}
\int V(x,z(t))\,\mathrm{d}t \approx -\frac{b}{v_z} C_3
                             \left[(fd/2-x)^{-3} + (fd/2+x)^{-3}\right].
\end{equation}

In the quantum model the potential of Eq.~\ref{eq:potint} leads to a phase
variation of the de Broglie waves passing at different position $x$ through an
individual slit of the second grating. In the classical model we expect the
potential gradient to lead to a position-dependent force which ultimately makes
an individual slit appear as a mini-lens with a velocity-dependent focal length.
By a simple geometrical argument, one expects maximal visibility for
$f_v\approx-L_1$, for our parameters  at velocities around 1000\,m/s. Indeed,
our classical simulation shows a single broad maximum of 18\,\% height in this
range. In the experimentally accessible velocity range, it predicts much lower
visibilities than our experimental data (dash-dotted line in
Fig.~\ref{fig:vdepend}). On the other hand, the measured functional dependence
of fringe visibility on velocity corresponds reasonably well to the numerical
quantum model with van der Waals interaction included (solid line in
Fig.~\ref{fig:vdepend}): the position of the single maximum coincides perfectly,
and the maximum experimental contrast attains nearly what is expected from
theory. Both the theoretical and the experimental curves show a single maximum
at 115\,m/s and are therefore asymmetrical with respect to the criterion
$L_1=L_T$
--- a consequence of the complex transmission function of the combined
absorptive and phase grating. The remaining discrepancy is not astonishing, having
in mind the rather coarse approximations underlying Eq.~\ref{eq:potint}, such as
the neglect of edge effects, and the inevitable experimental imperfections.

Talbot-Lau interferometers allow various interferometric measurements, especially
those of inertial forces or of decoherence effects, which do not require spatially
separated beams. As a first example, we have investigated the gravitational phase
shift in our interferometer by tilting the optical table with the whole
experimental setup by few milliradians. As expected \cite{Oberthaler1996a} the
interference fringes are shifted by a phase of $\Delta\varphi_g=2\pi L_1^2
g\alpha/(d v^2)=0.2\,$rad per mrad of table inclination $\alpha$. This means that
the gravitational acceleration component in the interferometer plane, which is on
the order of $10^{-3}g$, can be measured within a short integration time.


 Concluding, we have demonstrated for the first time an interferometer for
massive molecules that are internally in a highly excited thermal state. The
fact that the experimental visibility is not significantly lower than the
theoretical expectation shows that decoherence by emission of black-body
radiation does not play a significant role in this experiment. However, there is
a chance to study decoherence in future experiments involving laser heating of
the fullerene molecules.

The successful implementation of this near-field interferometer paves the way
towards interference of even more massive objects. Talbot-Lau interferometry has a
favorable scaling behaviour. If one leaves the longitudinal length scale
$L_T=d^2/\lambda$ unchanged, a reduction of the grating period by a factor of four
already allows the observation of quantum interference for a 16 times shorter de
Broglie wavelength
--- that means 16 times heavier molecules, if one assumes a beam of equal
velocity, or even 256 times with effusive characteristics at the same temperature.
However, the present experiment teaches us that one has to take into account the
effect of the van der Waals interaction. With decreasing grating period the
influence of the $r^{-3}$ potential increases strongly. In numerical simulations,
the resulting contributions of high diffraction orders at the second grating
restrict the self-imaging phenomenon and therefore the fringe visibility to
extremely narrow velocity ranges, e.~g.\ $\Delta v / v \approx 0.01$ for
$m=16\:m_\mathrm{C70}$ and $d=d_\mathrm{C70}/4$, assuming both an unchanged
velocity and an unchanged $C_3$ constant for simplicity. For experimentally
accessible velocity distributions, the inclusion of the van der Waals interaction
gives therefore a more stringent limit to interferometer scalability than
previously discussed by Schmiedmayer et~al.\ \cite{Berman1997a}.

For an experimental implementation, it seems therefore more promising to implement
the second grating as an optical phase grating like the one which we have recently
demonstrated for fullerenes \cite{Nairz01a}. Exploiting the position-dependent
ionization and fragmentation probability of molecules in a strong standing light
wave, one could also replace the first and third gratings by laser light, thereby
eliminating the fragile free-standing microstructures~\cite{Dubetsky1999b}.

\begin{acknowledgments}
We acknowledge help in the design of the experiment by Gerbrand van der Zouw and
William Case. This work has been supported by the European TMR network, contract
No.\ ERBFMRXCT960002, by the Austrian Science Foundation (FWF), within the
projects F1505 and START Y177, and by the Marie Curie Fellowship
HPMF-CT-2000-00797 of the European Community (B.~B.).
\end{acknowledgments}


\begin{thebibliography}{20}
\expandafter\ifx\csname natexlab\endcsname\relax\def\natexlab#1{#1}\fi
\expandafter\ifx\csname bibnamefont\endcsname\relax
  \def\bibnamefont#1{#1}\fi
\expandafter\ifx\csname bibfnamefont\endcsname\relax
  \def\bibfnamefont#1{#1}\fi
\expandafter\ifx\csname citenamefont\endcsname\relax
  \def\citenamefont#1{#1}\fi
\expandafter\ifx\csname url\endcsname\relax
  \def\url#1{\texttt{#1}}\fi
\expandafter\ifx\csname urlprefix\endcsname\relax\def\urlprefix{URL }\fi
\providecommand{\bibinfo}[2]{#2} \providecommand{\eprint}[2][]{\url{#2}}

\bibitem[{\citenamefont{Aspect and (Eds.)}(2001)}]{Aspect2001a}
Special issue edited by
\bibinfo{author}{\bibfnamefont{A.}~\bibnamefont{Aspect}} \bibnamefont{and}
  \bibinfo{author}{\bibfnamefont{J.}~\bibnamefont{Dalibard}},
  \bibinfo{journal}{Comptes rendus de l'Acad{\'e}mie des sciences},
  \textbf{{t.~2, S\'erie IV}}, No.~4 (\bibinfo{year}{2001}).

\bibitem[{\citenamefont{Berman}(1997)}]{Berman1997a}
  \emph{\bibinfo{title}{Atom Interferometry}},
  edited by \bibinfo{editor}{\bibfnamefont{P.~R.} \bibnamefont{Berman}}
   (\bibinfo{publisher}{Academic Press},
   \bibinfo{address}{San Diego}, \bibinfo{year}{1997}).

\bibitem[{\citenamefont{Rauch and Werner}(2000)}]{Rauch2000a}
\bibinfo{author}{\bibfnamefont{H.}~\bibnamefont{Rauch}} \bibnamefont{and}
  \bibinfo{author}{\bibfnamefont{S.~A.} \bibnamefont{Werner}},
  \emph{\bibinfo{title}{Neutron interferometry}} (\bibinfo{publisher}{Oxford
  University Press},
  \bibinfo{address}{New York},
  \bibinfo{year}{2000}).

\bibitem[{\citenamefont{Gustavson et~al.}(1997)\citenamefont{Gustavson, Bouyer,
  and Kasevich}}]{Gustavson1997a}
\bibinfo{author}{\bibfnamefont{T.~L.} \bibnamefont{Gustavson}},
  \bibinfo{author}{\bibfnamefont{P.}~\bibnamefont{Bouyer}}, \bibnamefont{and}
  \bibinfo{author}{\bibfnamefont{M.~A.} \bibnamefont{Kasevich}},
  \bibinfo{journal}{Phys. Rev. Lett.} \textbf{\bibinfo{volume}{78}},
  \bibinfo{pages}{2046} (\bibinfo{year}{1997}).

\bibitem[{\citenamefont{Weiss et~al.}(1993)\citenamefont{Weiss, Young, and
  Chu}}]{Weiss1993a}
\bibinfo{author}{\bibfnamefont{D.~S.} \bibnamefont{Weiss}},
  \bibinfo{author}{\bibfnamefont{B.~C.} \bibnamefont{Young}}, \bibnamefont{and}
  \bibinfo{author}{\bibfnamefont{S.}~\bibnamefont{Chu}},
  \bibinfo{journal}{Phys. Rev. Lett.} \textbf{\bibinfo{volume}{70}},
  \bibinfo{pages}{2706} (\bibinfo{year}{1993}).

\bibitem[{\citenamefont{Arndt et~al.}(2002)\citenamefont{Arndt, Nairz, and
  Zeilinger}}]{ArndtBellbook}
\bibinfo{author}{\bibfnamefont{M.}~\bibnamefont{Arndt}},
  \bibinfo{author}{\bibfnamefont{O.}~\bibnamefont{Nairz}}, \bibnamefont{and}
  \bibinfo{author}{\bibfnamefont{A.}~\bibnamefont{Zeilinger}}, in
  \emph{\bibinfo{booktitle}{Quantum $[$Un$]$Speakables}}, edited by
  \bibinfo{editor}{\bibfnamefont{R.}~\bibnamefont{Bertlmann}} \bibnamefont{and}
  \bibinfo{editor}{\bibfnamefont{A.}~\bibnamefont{Zeilinger}}
  (\bibinfo{publisher}{Springer},
  \bibinfo{address}{New York},
  \bibinfo{year}{2002}).

\bibitem[{\citenamefont{Bord{\'e} et~al.}(1994)\citenamefont{Bord{\'e},
  Courtier, Burck, Goncharov, and Gorlicki}}]{Borde1994a}
\bibinfo{author}{\bibfnamefont{C.}~\bibnamefont{Bord{\'e}}},
  \bibinfo{author}{\bibfnamefont{N.}~\bibnamefont{Courtier}},
  \bibinfo{author}{\bibfnamefont{F.~D.} \bibnamefont{Burck}},
  \bibinfo{author}{\bibfnamefont{A.}~\bibnamefont{Goncharov}},
  \bibnamefont{and} \bibinfo{author}{\bibfnamefont{M.}~\bibnamefont{Gorlicki}},
  \bibinfo{journal}{Phys. Lett. A} \textbf{\bibinfo{volume}{188}},
  \bibinfo{pages}{187} (\bibinfo{year}{1994}).

\bibitem[{\citenamefont{Chapman
  et~al.}(1995{\natexlab{a}})\citenamefont{Chapman, Ekstrom, Hammond,
  Rubenstein, Schmiedmayer, Wehinger, and Pritchard}}]{Chapman1995b}
\bibinfo{author}{\bibfnamefont{M.~S.} \bibnamefont{Chapman}},
  \bibinfo{author}{\bibfnamefont{C.~R.} \bibnamefont{Ekstrom}},
  \bibinfo{author}{\bibfnamefont{T.~D.} \bibnamefont{Hammond}},
  \bibinfo{author}{\bibfnamefont{R.~A.} \bibnamefont{Rubenstein}},
  \bibinfo{author}{\bibfnamefont{J.}~\bibnamefont{Schmiedmayer}},
  \bibinfo{author}{\bibfnamefont{S.}~\bibnamefont{Wehinger}}, \bibnamefont{and}
  \bibinfo{author}{\bibfnamefont{D.~E.} \bibnamefont{Pritchard}},
  \bibinfo{journal}{Phys. Rev. Lett.} \textbf{\bibinfo{volume}{74}},
  \bibinfo{pages}{4783} (\bibinfo{year}{1995}{\natexlab{a}}).

\bibitem[{\citenamefont{Clauser and Li}(1994)}]{Clauser1994a}
\bibinfo{author}{\bibfnamefont{J.~F.} \bibnamefont{Clauser}} \bibnamefont{and}
  \bibinfo{author}{\bibfnamefont{S.}~\bibnamefont{Li}}, \bibinfo{journal}{Phys.
  Rev. A} \textbf{\bibinfo{volume}{49}}, \bibinfo{pages}{R2213}
  (\bibinfo{year}{1994}).

\bibitem[{\citenamefont{Deng et~al.}(1999)\citenamefont{Deng, Hagley,
  Denschlag, Simsarian, Edwards, Clark, Helmerson, Rolston, and
  Phillips}}]{Deng1999b}
\bibinfo{author}{\bibfnamefont{L.}~\bibnamefont{Deng}},
  \bibinfo{author}{\bibfnamefont{E.~W.} \bibnamefont{Hagley}},
  \bibinfo{author}{\bibfnamefont{J.}~\bibnamefont{Denschlag}},
  \bibinfo{author}{\bibfnamefont{J.~E.} \bibnamefont{Simsarian}},
  \bibinfo{author}{\bibfnamefont{M.}~\bibnamefont{Edwards}},
  \bibinfo{author}{\bibfnamefont{C.~W.} \bibnamefont{Clark}},
  \bibinfo{author}{\bibfnamefont{K.}~\bibnamefont{Helmerson}},
  \bibinfo{author}{\bibfnamefont{S.~L.} \bibnamefont{Rolston}},
  \bibnamefont{and} \bibinfo{author}{\bibfnamefont{W.~D.}
  \bibnamefont{Phillips}}, \bibinfo{journal}{Phys. Rev. Lett.}
  \textbf{\bibinfo{volume}{83}}, \bibinfo{pages}{5407} (\bibinfo{year}{1999}).

\bibitem[{\citenamefont{Clauser}(1997)}]{Clauser1997a}
\bibinfo{author}{\bibfnamefont{J.}~\bibnamefont{Clauser}}, in
  \emph{\bibinfo{booktitle}{Experimental Metaphysics}}, edited by
  \bibinfo{editor}{\bibfnamefont{R.}~\bibnamefont{Cohen}} \bibnamefont{et~al.}
  (\bibinfo{publisher}{Kluwer},
  \bibinfo{address}{Dordrecht},
  \bibinfo{year}{1997}).

\bibitem[{\citenamefont{Arndt et~al.}(1999)\citenamefont{Arndt, Nairz,
  Voss-Andreae, Keller, der Zouw, and Zeilinger}}]{Arndt1999a}
\bibinfo{author}{\bibfnamefont{M.}~\bibnamefont{Arndt}},
  \bibinfo{author}{\bibfnamefont{O.}~\bibnamefont{Nairz}},
  \bibinfo{author}{\bibfnamefont{J.}~\bibnamefont{Voss-Andreae}},
  \bibinfo{author}{\bibfnamefont{C.}~\bibnamefont{Keller}},
  \bibinfo{author}{\bibfnamefont{G.~V.} \bibnamefont{der Zouw}},
  \bibnamefont{and}
  \bibinfo{author}{\bibfnamefont{A.}~\bibnamefont{Zeilinger}},
  \bibinfo{journal}{Nature} \textbf{\bibinfo{volume}{401}},
  \bibinfo{pages}{680} (\bibinfo{year}{1999}).

\bibitem[{\citenamefont{Nairz et~al.}(2001)\citenamefont{Nairz, Brezger, Arndt,
  and Zeilinger}}]{Nairz01a}
\bibinfo{author}{\bibfnamefont{O.}~\bibnamefont{Nairz}},
  \bibinfo{author}{\bibfnamefont{B.}~\bibnamefont{Brezger}},
  \bibinfo{author}{\bibfnamefont{M.}~\bibnamefont{Arndt}}, \bibnamefont{and}
  \bibinfo{author}{\bibfnamefont{A.}~\bibnamefont{Zeilinger}},
  \bibinfo{journal}{Phys. Rev. Lett.} \textbf{\bibinfo{volume}{87}},
  \bibinfo{pages}{160401} (\bibinfo{year}{2001}).

\bibitem[{\citenamefont{Talbot}(1836)}]{Talbot1836}
\bibinfo{author}{\bibfnamefont{H.~F.} \bibnamefont{Talbot}},
  \bibinfo{journal}{Philos. Mag.} \textbf{\bibinfo{volume}{9}},
  \bibinfo{pages}{401} (\bibinfo{year}{1836}).

\bibitem[{\citenamefont{Chapman
  et~al.}(1995{\natexlab{b}})\citenamefont{Chapman, Ekstrom, Hammond,
  Schmiedmayer, Tannian, Wehinger, and Pritchard}}]{Chapman1995a}
\bibinfo{author}{\bibfnamefont{M.~S.} \bibnamefont{Chapman}},
  \bibinfo{author}{\bibfnamefont{C.~R.} \bibnamefont{Ekstrom}},
  \bibinfo{author}{\bibfnamefont{T.~D.} \bibnamefont{Hammond}},
  \bibinfo{author}{\bibfnamefont{J.}~\bibnamefont{Schmiedmayer}},
  \bibinfo{author}{\bibfnamefont{B.~E.} \bibnamefont{Tannian}},
  \bibinfo{author}{\bibfnamefont{S.}~\bibnamefont{Wehinger}}, \bibnamefont{and}
  \bibinfo{author}{\bibfnamefont{D.~E.} \bibnamefont{Pritchard}},
  \bibinfo{journal}{Phys. Rev. A} \textbf{\bibinfo{volume}{51}},
  \bibinfo{pages}{R14} (\bibinfo{year}{1995}{\natexlab{b}}).

\bibitem[{\citenamefont{Patorski}(1989)}]{Patorski1989a}
\bibinfo{author}{\bibfnamefont{K.}~\bibnamefont{Patorski}},
  \bibinfo{journal}{Prog. Opt.} \textbf{\bibinfo{volume}{27}},
  \bibinfo{pages}{1} (\bibinfo{year}{1989}).

\bibitem{footnote1}
 These specific values refer to Fig.~\ref{fig:vdepend}.

\bibitem[{\citenamefont{Oberthaler et~al.}(1996)\citenamefont{Oberthaler,
  Bernet, Rasel, Schmiedmayer, and Zeilinger}}]{Oberthaler1996a}
\bibinfo{author}{\bibfnamefont{M.~K.} \bibnamefont{Oberthaler}},
  \bibinfo{author}{\bibfnamefont{S.}~\bibnamefont{Bernet}},
  \bibinfo{author}{\bibfnamefont{E.~M.} \bibnamefont{Rasel}},
  \bibinfo{author}{\bibfnamefont{J.}~\bibnamefont{Schmiedmayer}},
  \bibnamefont{and}
  \bibinfo{author}{\bibfnamefont{A.}~\bibnamefont{Zeilinger}},
  \bibinfo{journal}{Phys. Rev. A} \textbf{\bibinfo{volume}{54}},
  \bibinfo{pages}{3165} (\bibinfo{year}{1996}).

\bibitem[{\citenamefont{Grisenti et~al.}(1999)\citenamefont{Grisenti,
  Sch{\"o}llkopf, Toennies, Hegerfeldt, and K{\"o}hler}}]{Grisenti1999a}
\bibinfo{author}{\bibfnamefont{R.~E.} \bibnamefont{Grisenti}},
  \bibinfo{author}{\bibfnamefont{W.}~\bibnamefont{Sch{\"o}llkopf}},
  \bibinfo{author}{\bibfnamefont{J.~P.} \bibnamefont{Toennies}},
  \bibinfo{author}{\bibfnamefont{G.~C.} \bibnamefont{Hegerfeldt}},
  \bibnamefont{and}
  \bibinfo{author}{\bibfnamefont{T.}~\bibnamefont{K{\"o}hler}},
  \bibinfo{journal}{Phys. Rev. Lett.} \textbf{\bibinfo{volume}{83}},
  \bibinfo{pages}{1755} (\bibinfo{year}{1999}).

\bibitem[{\citenamefont{Vidali et~al.}(1991)\citenamefont{Vidali, Ihm, Kim, and
  Cole}}]{Vidali1991a}
\bibinfo{author}{\bibfnamefont{G.}~\bibnamefont{Vidali}},
  \bibinfo{author}{\bibfnamefont{G.}~\bibnamefont{Ihm}},
  \bibinfo{author}{\bibfnamefont{H.~Y.} \bibnamefont{Kim}}, \bibnamefont{and}
  \bibinfo{author}{\bibfnamefont{M.~W.} \bibnamefont{Cole}},
  \bibinfo{journal}{Surf. Sci. Rep.} \textbf{\bibinfo{volume}{12}},
  \bibinfo{pages}{133} (\bibinfo{year}{1991}).

\bibitem[{\citenamefont{Dresselhaus et~al.}(1998)\citenamefont{Dresselhaus,
  Dresselhaus, and Eklund}}]{Dresselhaus1998a}
\bibinfo{author}{\bibfnamefont{M.~S.} \bibnamefont{Dresselhaus}},
  \bibinfo{author}{\bibfnamefont{G.}~\bibnamefont{Dresselhaus}},
  \bibnamefont{and} \bibinfo{author}{\bibfnamefont{P.~C.}
  \bibnamefont{Eklund}}, \emph{\bibinfo{title}{Science of Fullerenes and Carbon
  Nanotubes}} (\bibinfo{publisher}{Academic Press}, \bibinfo{address}{San Diego},
  \bibinfo{year}{1998}), \bibinfo{edition}{2nd} ed.

\bibitem[{\citenamefont{Dubetsky and Berman}(1999)}]{Dubetsky1999b}
Combinations of absorptive and phase gratings have been discussed for the quite
different case of atoms by
\bibinfo{author}{\bibfnamefont{B.}~\bibnamefont{Dubetsky}} \bibnamefont{and}
  \bibinfo{author}{\bibfnamefont{P.~R.} \bibnamefont{Berman}}
  [\bibinfo{journal}{Phys. Rev. A} \textbf{\bibinfo{volume}{59}},
  \bibinfo{pages}{2269} (\bibinfo{year}{1999})].

\end{thebibliography}


\end{document}